\documentclass[11pt]{article}
\usepackage{moriond}

\def\be{\begin{equation}}
\def\ee{\end{equation}}
\def\bea{\begin{eqnarray}}
\def\eea{\end{eqnarray}}

\newcommand{\Photo}{}

\begin{document}
\hfill{DFPD-13/TH/18}
\vspace*{4cm}
\title{THEORY SUMMARY}

\author{FABIO ZWIRNER}

\address{Dipartimento di Fisica ed Astronomia `G.~Galilei', Universit\`a degli Studi di Padova, \\
and Istituto Nazionale di Fisica Nucleare, Sezione di Padova, \\ 
Via Marzolo 8,  35131 Padova, Italy}

\maketitle\abstracts{Talk given at the XLVIII Rencontres de Moriond, Electroweak Interactions and Unified Theories, La Thuile, Italy, 2-9 March 2013, to appear in the Proceedings.}

\newpage

\section{Preamble}

I thank the Organizing Committee for assigning me the honour and the challenge of this Theory Summary. 
%A theorist in the Organizing Committee recommended me to give a broad review, without indulging in describing my own work: you will see from the list of references that I took his recommendation literally. 
My talk comes after an outstanding Experimental Summary~\cite{sphicas}, which gave full justice to the unusual richness and importance of the experimental results presented. 
The theory program of the meeting was also very rich, although perhaps less systematic: I counted 26 `regular' theory talks and 10 `short' theory talks by young scientists. 
It would be impossible (and boring) to give a fair account of all of them.
I will focus instead on presenting my view of the status and the prospects of the field, as emerging also from the theory talks at this meeting, but without reviewing the latter in a systematic way: I apologise in advance for the many omissions.
The four headlines of my opening slide are collected in Table~\ref{tab:headlines}.

\begin{table}[hbt]
\caption{The four headlines of the Electroweak Rencontres de Moriond 2013 according to the speaker.}
\label{tab:headlines}
\vspace{0,3cm}
\begin{center}
\begin{tabular}{|c|c|}
\hline
& \\
HEP-EX is on the move
&
The triumph of the SM
\\ 
& \\
\hline
& \\
Where is BSM physics?
&
Naturalness challenged
\\
& \\
\hline
\end{tabular}
\end{center}
\end{table}
\vspace{0.4cm}

The meaning of the top left headline should be clear from the Experimental Summary:  the impressive flow of LHC results on a very wide spectrum of physics topics, the last results from the Tevatron and the B-factories, many results from precision experiments, neutrino experiments, astroparticle experiments, the results from the Planck mission (to be released soon after the end of this meeting) are giving strong new input, although not necessarily the desired one, to particle physics, and are greatly helping the theorists who want to be helped by data, in particular those interested in Beyond-the-Standard-Model (BSM) physics, to re-focus their goals.

\section{The triumph of the Standard Model}

The Standard Model (SM) of strong and electroweak interactions, effectively coupled to gravity, quantitatively describes most observations. 
Neutrino oscillations, although not accounted for by the very minimal formulation of the SM, so far call only for minor modifications (see subsection 3.1). 
Much stronger modifications are probably~\footnote{There is also a radical point of view~\cite{shapo}, according to which the SM with the addition of right-handed neutrinos, one in the keV region and the other two in the GeV region, is everything we need up to the Planck scale to explain Dark Matter, Inflation and Baryogenesis.} required to account for phenomena having to do with gravitation, astrophysics and cosmology, such as Dark Matter, Dark Energy, Inflation and Baryogenesis.
The above statements were true last year, they are still true today: what has then changed in our understanding?

Let me start from the big question of particle physics to which we are finding answers now: the problem of symmetry breaking in the electroweak theory. 
There are many other unanswered big questions in particle physics, equally or even more important: what makes the above question special is the fact that we have started answering it experimentally, and we are likely to get a reasonably complete picture on the time scale of the present generation of physicists and experiments.  

The part of the minimal SM Lagrangian involving only the gauge bosons and the fermions, schematically written as
\begin{equation}
\label{smLGF}
{\cal L}_{GF} =
-{1 \over 4} F_{\mu \nu}  F^{\mu \nu} 
+
i \, \overline{\Psi} D \!\!\!\! \slash \  \Psi \, , 
\end{equation}
has the local (gauge) symmetry $G_{loc} = SU(3)_C \times SU(2)_L \times U(1)_Y$ and an independent global symmetry  $G_{gl} = [SU(3)]^5 \times [U(1)]^4$. 
The part containing also the complex scalar doublet $\phi$, schematically written as
\begin{equation}
\label{smlSY}
{\cal L}_{SY} = 
\left( D_{\mu} \phi \right)^{\dagger}
\left( D^{\mu} \phi \right)
-  \mu^2 \phi^{\dagger} \phi
- \lambda \left( \phi^{\dagger} \phi  \right)^2 
+ \left( \overline{\Psi}_i \ Y_{ij} \ \Psi_j \ \phi +  {\rm h.c.} \right) \, ,
\end{equation}
has the twofold role of spontaneously breaking the gauge symmetry to $H_{loc} = SU(3)_C \times U(1)_{em}$ and the global symmetry to  $H_{gl} = [U(1)]^4$, associated with the baryon number $B$ and the three partial lepton numbers $(L_e,L_\mu,L_\tau)$. 

The present message from experiment to theory is clear: this minimal, weakly-coupled SM implementation of the Brout-Englert-Higgs mechanism,  with  a single ÒelementaryÓ scalar doublet $\phi$ in the Lagrangian, the Cabibbo-Kobayashi-Maskawa (CKM) description of flavour change and CP violation, and a generalised Glashow-Iliopoulos-Maiani (GIM) mechanism at work for the suppression of flavour-changing neutral currents, works {\em far beyond most expectations}. 
This message is not entirely new, and has been receiving increasing support over the past decades from the unsuccessful direct searches for non-SM particles and from the successful precision tests of the SM  at LEP, at the Tevatron, and at the various experiments at the intensity frontier.  
But now we have new and more powerful experimental milestones in support of the SM. 
The outstanding one is undoubtedly the discovery of a new particle, to be called {\em The Boson} in the following, whose properties are so far compatible with those of the SM scalar: we are finally ready to change our introductions to particle physics courses, and tell our students that there are not four but five fundamental forces in Nature, mediated by spin-0, spin-1 and spin-2 bosons!  
We have also new precise tests of the SM in the realm of flavour physics and CP violation, all passed with flying colours. 
And the bounds on hypothetical non-SM particles are much stronger than before, in particular the new bounds from ATLAS and CMS extend well above 1 TeV for all those particles that have sizeable couplings to quarks and gluons and can give rise to detectable signatures in the challenging LHC environment.

\subsection{Flavour physics}

In reviewing the enduring triumph of the SM, I will start right away from flavour physics, since it was the subject of the opening talk of the meeting~\cite{maiani}. 
After reminding us of the history of the GIM mechanism, Maiani reviewed how its generalisation to the SM has been more and more precisely tested over the years, in a long series of measurements of increasing variety and precision. 
Today, precision tests of the SM in the flavour sector put very stringent bounds on possible New Physics, much above the reach of the LHC direct searches if such New Physics couples generically to the SM. 
It is a harder and harder challenge for theorists to come up with motivated models that predict new states within the kinematical reach of the LHC, but can also evade the flavour constraints with a minimum of theoretical dignity. 
However, theorists are not easily discouraged and some of them are trying to meet the challenge.

\begin{figure}
\begin{minipage}{0.48\linewidth}
\centerline{\includegraphics[width=0.8\linewidth]{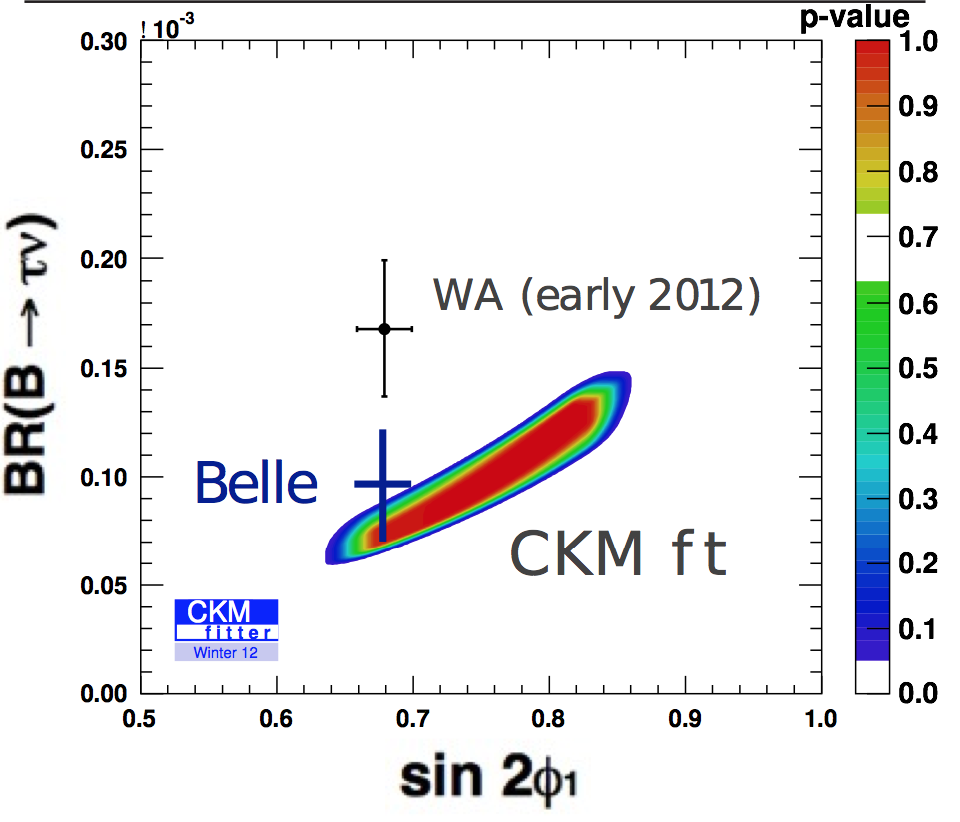}}
\end{minipage}
\hfill
\begin{minipage}{0.48\linewidth}
\centerline{\includegraphics[width=0.9\linewidth]{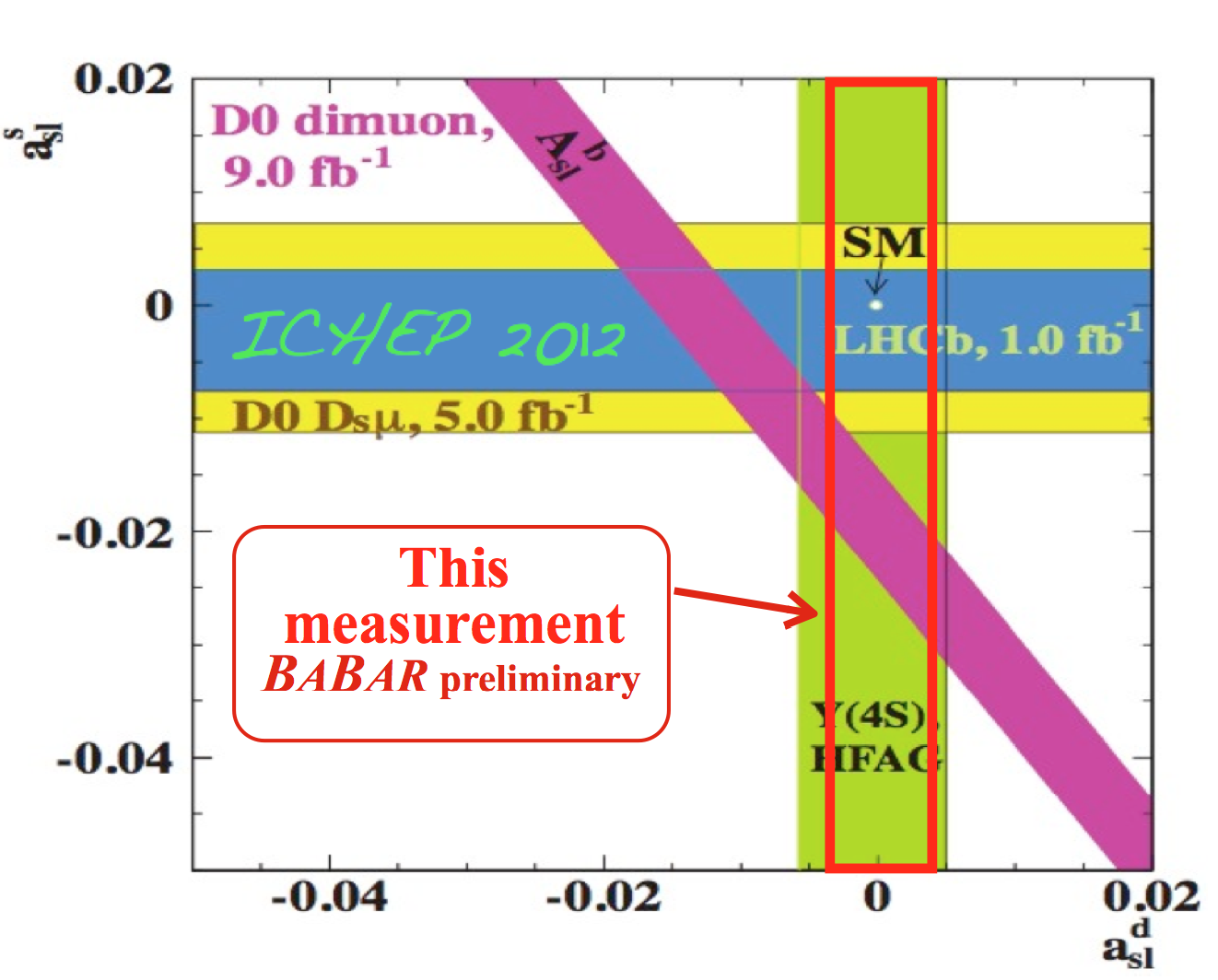}}
\end{minipage}
\begin{minipage}{0.48\linewidth}
\centerline{\includegraphics[width=0.9\linewidth]{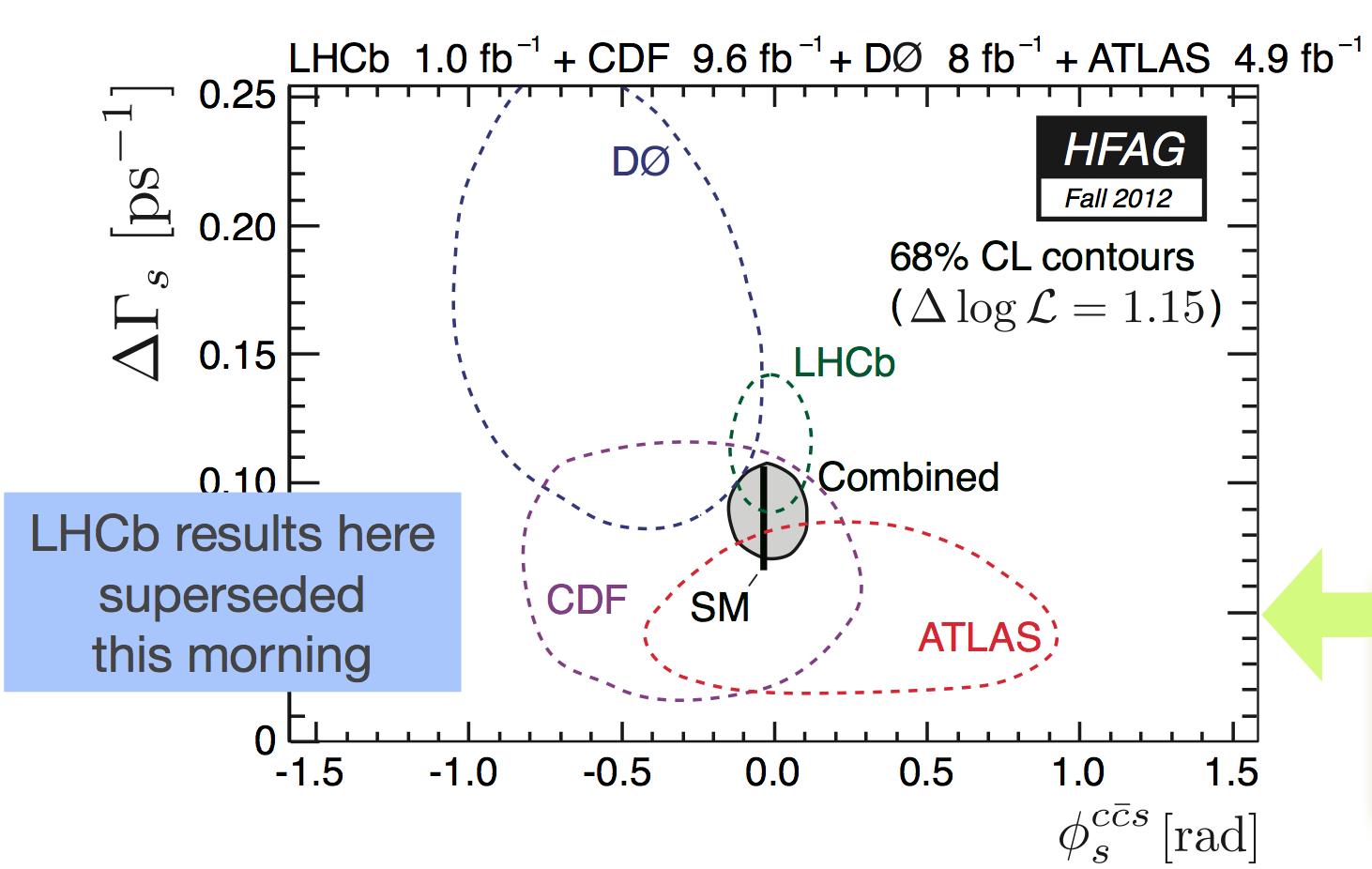}}
\end{minipage}
\hfill
\begin{minipage}{0.48\linewidth}
\centerline{\includegraphics[width=0.8\linewidth]{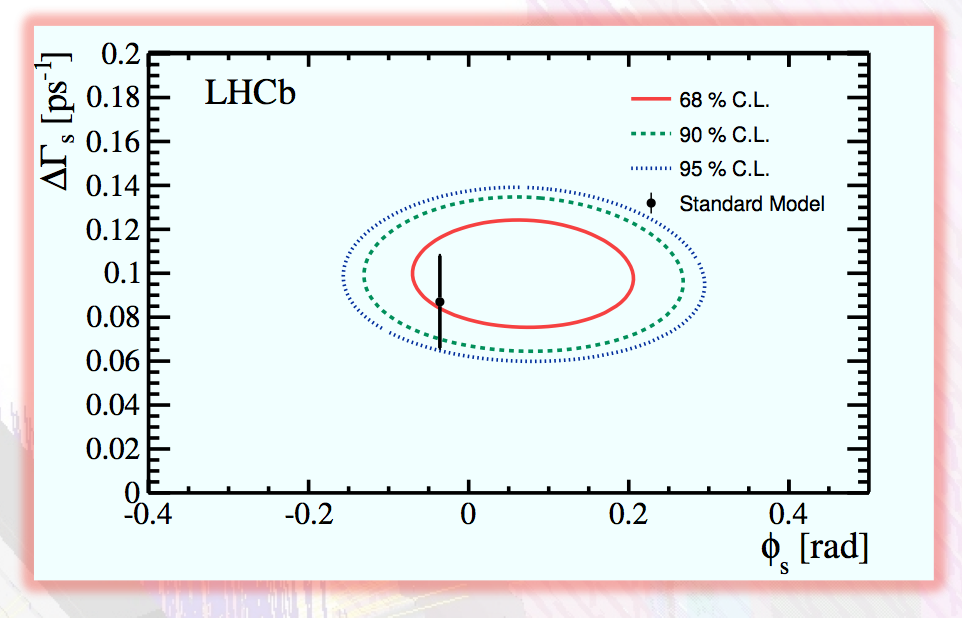}}
\end{minipage}
\caption{Some of the many small tensions in B physics now fading away, thanks to the new results from Belle on $B \to \tau \nu$,  from Babar on CP violation in $B^0$ mixing and from ATLAS and LHCb on $B_s \rightarrow J /  \psi \, \phi$.}
\label{fig:btensions1}
\end{figure}

On the experimental side, many small tensions, which raised some hopes and triggered many theoretical papers in the last years, are now fading away thanks to new results: some examples are given in Fig.~\ref{fig:btensions1}, taken from presentations by Belle~\cite{huschle} (top left),  BaBaR~\cite{akar} (top right), ATLAS~\cite{spagnolo} (bottom left)  and LHCb~\cite{benson} (bottom right).
There were also some theory-motivated hopes that have not materialised (so far): outstanding examples are the new MEG bound~\cite{ootani} $ BR (\mu \rightarrow e \, \gamma)   < 5.7 \times 10^{-13}$  $(90\% \, c.l.)$  and the $3.5 \sigma$ evidence from LHCb~\cite{sarti} for the rare decay $B_s \rightarrow \mu^+ \mu^-$, at a rate in agreement with the SM prediction. 
The latter result is expected to be updated, not only by LHCb but also by CMS, at the Summer conferences. 
Other small tensions of the past are losing significance.
The CDF top forward-backward asymmetry has now been replaced by a more sophisticated study of the full angular distribution~\cite{wilson}, and in a projection onto Legendre polynomials the excess in the linear term coefficient is only $2.1 \sigma$ away from the NLO SM prediction. 
These results start being tested also at the LHC~\cite{battilana}.
The quantity $\Delta A_{CP} = A_{CP}(KK) - A_{CP}(\pi\pi)$ in D decays, found at face value to be $3.5 \sigma$ away from the SM prediction in LHCb, has been followed by re-assessments of the theoretical uncertainties,  and an experimental update is imminent~\cite{hampson}.

On the theory side, precision flavour physics calls for state-of-the-art flavour phenomenology. 
A hot example discussed at this meeting are the $b \rightarrow s l^+ l^-$ transitions, for example $B \rightarrow K^* \mu^+ \mu^-$, since the Tevatron experiments and LHCb have entered the precision era and started measuring angular distributions. 
Hiller discussed the predictions for the angular distributions at low hadronic recoil and the extraction of hadronic form factors from data~\cite{hiller}. 
Virto discussed how to extract constraints on flavour effective operators with theoretically `clean' observables, and how to deduce in turn new strong constraints on semileptonic and radiative operators~\cite{virto}.

Precision flavour physics keeps calling also for hard SM theory, for example for state-of-the-art lattice calculations. 
A very challenging problem is the derivation of the $\Delta I = 1/2$ rule from QCD on the lattice.
This rule accounts for the fact that $\Gamma(K_s \rightarrow \pi^+ \pi^-)/ \Gamma(K^+ \rightarrow \pi^+ \pi^0) \simeq 670$, corresponding to $(Re A_0)/(Re A_2) \simeq 22.5$, with important implications for the calculation of $\epsilon^\prime/\epsilon$.  
We heard from Soni~\cite{soni} that using domain-wall quarks, full QCD rather than chiral perturbation theory and the physical pion mass, a new strong cancellation was found, corresponding to a suppression factor of 3 to 4 in $Re A_2$. The calculation of $A_0$ needs more work but is under way, and there are hopes of substantial progress towards solving the long-standing $\Delta I = 1/2$ puzzle.

\subsection{The Boson}

The experimental results on The Boson, in particular the impressive progress in the study of its properties after its discovery in July 2012, have been extensively described in the Experimental Summary~\cite{sphicas},  and more updates are coming soon from Moriond QCD.
Without repeating what has already been said, I just add three personal comments:
\begin{enumerate}
\item
Once more, I would like to express my theorist's admiration for the marvellous achievements of the experimental colleagues in ATLAS and CMS. 
\item 
At this meeting, I was impressed by the new direct indications hinting at  SM-like couplings of The Boson to the $\tau$ lepton and to the $b$ quark. 
\item
To take seriously deviations from the properties of The Boson predicted by the SM, we should apply the same stringent standards as for discovery! 
It is then difficult to imagine a theoretical crisis in the SM interpretation of The Boson in less than 3 years.
\end{enumerate}

One of the dominating themes of the experimental talks on The Boson has been the following question: is it really a spin-0 CP-even particle? 
We heard that ATLAS and CMS are now testing the $J^{CP}$ properties of The Boson: this is very important as a consistency check and must be done. 
However, when reading in the experimental slides the number of standard deviations differentiating the $0^+$ hypothesis from the alternative ones, we should think for a moment of what these $\sigma$'s mean.
With the mass $M_H$ known by now with very good accuracy, no free parameter is left in the SM to describe all the production mechanisms and decay modes of The Boson.
Moreover, we are performing such computations in a renormalizable theory that gloriously passes all precision tests and can be safely extrapolated to (much) higher energies. 
This cannot be put on the same footing as the much more baroque attempts at describing the properties of an impostor with $J^{CP} \ne 0^+$! 
Technically, it is possible to write a suitable effective Lagrangian, but this adds many parameters and contributes to what I would call Ò\emph{theory $\sigma$'s}Ó.
Furthermore, making such effective Lagrangian compatible with both the observed properties of The Boson and all the electroweak precision tests may be non-trivial.

Assuming for now that The Boson is a CP-even spin-0 particle, the crucial question, now and for many years to come, is whether it can be identified with the SM scalar.  
There are several ways for The Boson to depart from the SM properties, as discussed in several theoretical talks~\cite{rzehak}$^, \,$\cite{carena}$^, \,$\cite{ellwanger}$^, \,$\cite{yamawaki}:
\begin{itemize}
\item
$H$ mixes with other spin-0 states, e.g. additional doublets and/or singlets.
\item
$H$ is the meson of a new strong force, kept light by its pseudo-Goldstone boson nature.
\item
$H$ decays into invisible particles.
\item
The loop diagrams controlling  $H$ production ($ggH$) and decay ($Hgg,H \gamma \gamma ,HZ \gamma$) are modified by new heavy particles that have so far escaped direct detection.
\end{itemize}

An important question, in view of the increasingly precise studies at the LHC and at possible future electron-positron colliders, is how to parametrize a non-SM scalar. 
This was the subject of several theoretical talks at this meeting~\cite{yamawaki}$^, \,$\cite{eboli}$^, \,$\cite{jenkins}$^, \,$\cite{azatov}$^, \,$\cite{gavela}$^, \,$\cite{merlo}.
We must find a compromise between simplicity and completeness, depending on the quality and type of the experimental data and on the purpose of the exercise.
For present book-keeping at the LHC, and under reasonable assumptions (spin-0, CP-even, no flavour-changing neutral-current couplings, no exotic decay channels), a simple but adequate parametrization consists in just rescaling the SM couplings to the massive vector bosons and the third generation fermions by suitable coefficients $\kappa_V$ ($V=W,Z$) and $\kappa_f$ ($f=t,b,\tau$), introducing additional coefficients $\kappa_{\gamma}$ and $\kappa_g$ for possible new physics contributions to the effective $H\gamma\gamma$ and $Hgg$ operators, respectively.
For testing models combining different sets of data, we should use suitable gauge-invariant effective Lagrangians built out of the SM scalar doublet $\phi$ and of the other SM fields. 
As discussed by Eboli~\cite{eboli} and Merlo~\cite{merlo}, the sets of operators encoding the leading new physics effects may be different depending on whether the dynamics at the origin of electroweak symmetry breaking is weakly or strongly interacting. 
In both cases, of course, it is important to identify  a suitable basis of independent operators.
Coming to more detailed issues, theoretically-motivated fitting strategies are being suggested by theorists to experimentalists~\cite{azatov}, and correlations between anomalous scalar couplings, electroweak precision tests and anomalous vector boson couplings are being identified~\cite{eboli}$^, \,$\cite{jenkins}.

A more general question, addressed in the talks by Strumia~\cite{strumia} and Elias Mir\'o~\cite{eliasmiro}, is what information can be extracted from the precisely measured mass of The Boson, $M_H = 125.6$ GeV with an uncertainty well below half a GeV in my personal back-of-the-envelope combination at the time of the meeting. 
\begin{figure}[htb]
\begin{minipage}{0.44\linewidth}
\centerline{\includegraphics[width=0.9\linewidth]{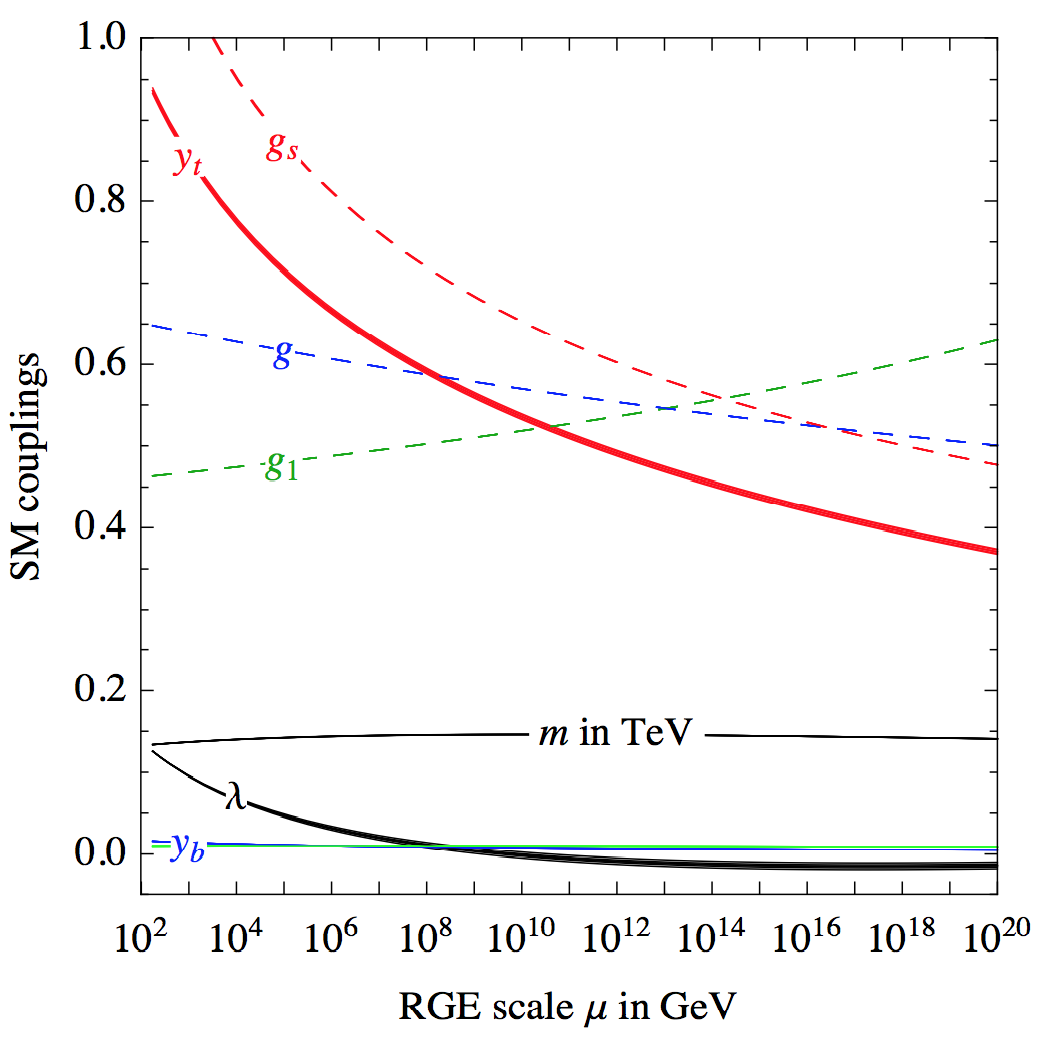}}
\end{minipage}
\hfill
\begin{minipage}{0.54\linewidth}
\centerline{\includegraphics[width=0.9\linewidth]{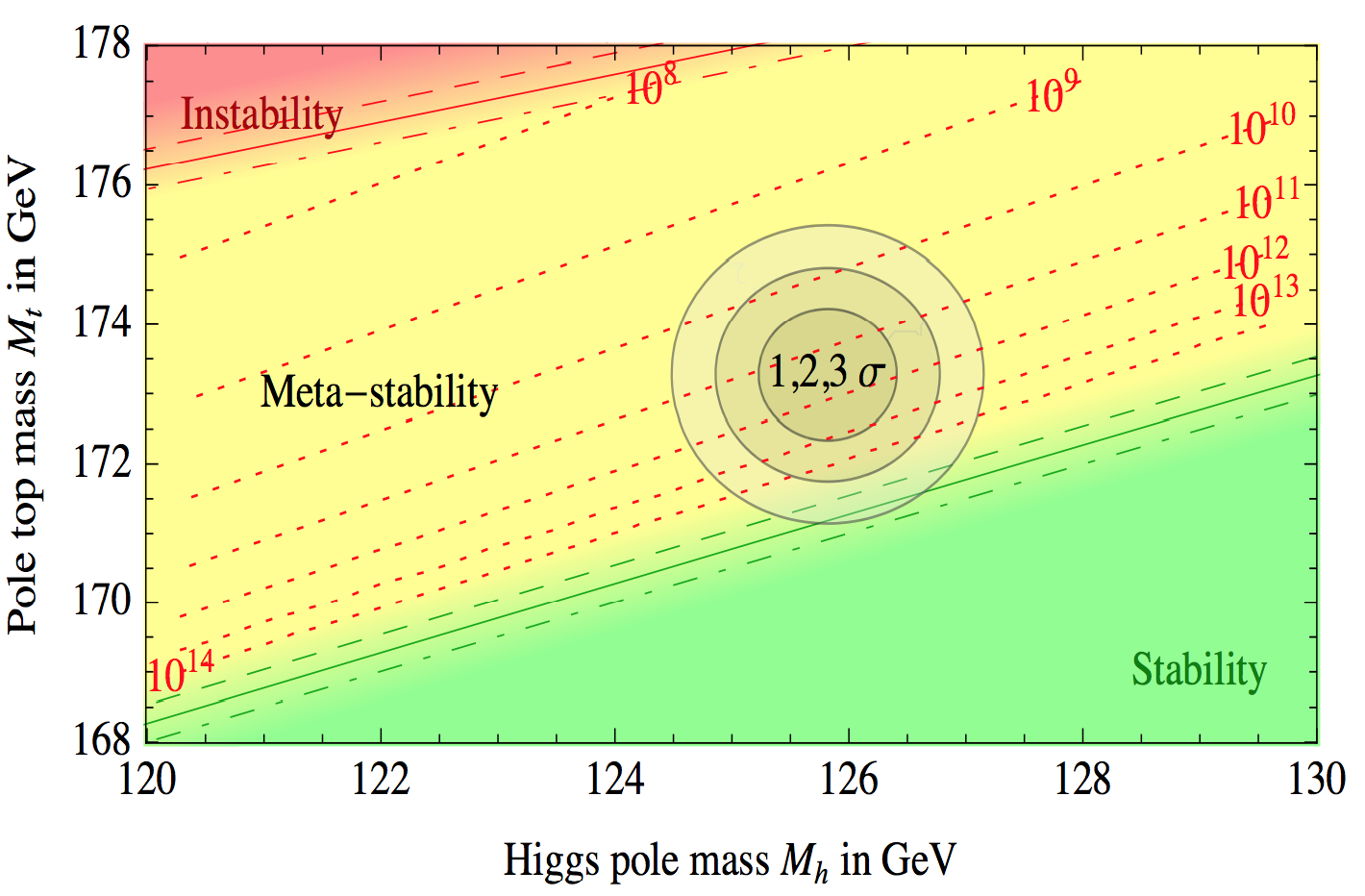}}
\end{minipage}
\caption{Left: renormalisation of the largest SM gauge and Yukawa couplings and of the SM scalar mass parameter $m$. Right: regions of stability, metastability and instability of the electroweak vacuum in the $(M_H,M_t)$ plane; contours at 1,2,3 $\sigma$ corresponding to the present knowledge of $\alpha_S$ are shown.}
\label{fig:higgstability}
\end{figure}
Indeed, values of $M_H$ different from the measured one could have implied that the SM cannot be extrapolated as such until the Planck scale $M_P$ of gravitational interactions, but at most up to a certain ultraviolet cutoff scale $\Lambda < M_P$. 
For sufficiently low values of $M_H$, say $M_H < 110 \, {\rm GeV}$, the SM effective potential would develop, besides the minimum corresponding to the experimental value of the electroweak scale, other minima with lower energy and much larger value of the scalar field. 
In first approximation, this corresponds to the fact that the SM effective scalar self-coupling, $\lambda(\mu)$, becomes negative at a scale $\mu < \Lambda$. 
However, the bound is slightly relaxed if we take into account that the correct electroweak vacuum can be metastable, as long as its lifetime is longer than the age of the universe.
Alternatively, for too large values of $M_H$, say $M_H > 175 \, {\rm GeV}$, the SM effective scalar self-coupling would develop a Landau pole at a scale $ \mu < \Lambda$.
Some intriguing consequences of the measured value of $M_H$ are shown~\cite{strumia} in Fig.~\ref{fig:higgstability}.
Taking into account the uncertainties in $M_H$, $M_t$ and $\alpha_S$, and assuming the validity of the SM up to the Planck scale, the electroweak vacuum is either absolutely stable or metastable, but with a lifetime many orders of magnitude larger than the age of the universe: as shown on the right-hand side of Fig.~\ref{fig:higgstability}, we are safe! 
Following the renormalisation group evolution of the SM couplings, we see from the left-hand side of  Fig.~\ref{fig:higgstability} that both the scalar self-coupling $\lambda (\mu)$ and $\beta_\lambda$, its rate of variation with respect to the scale $\mu$, become and stay quite close to zero at scales $\mu > 10^8 \, {\rm GeV}$.
What do we learn from this? 
\begin{itemize}
\item
The most important message is that there is nothing forcing us to extend the SM before a scale $\Lambda$  of order $10^{10}$~GeV or so (if we ignore naturalness, see subsection 3.3): the scales for right-handed neutrino masses in the simplest realisation of the see-saw mechanism and for the invisible axion as a dark matter candidate can be safely beyond.
\item
Is there some meaning in the near vanishing of $\lambda$, $\beta_\lambda$, $M_H/\Lambda$ at very high cutoff scales for the SM? Some theorists see this as the hint of a possible underlying spontaneously broken conformal symmetry, but no complete and self-consistent picture of how this could be concretely realised is available. My forecast, however, is that the theoretical efforts in this direction will increase.    
\item
In my opinion, more precise renormalization group calculations and more precise measurements of $M_t$, $M_H$ and $\alpha_S$ will become important only when and if threshold effects at the cutoff scale $\Lambda$ will be calculable in the framework of a well-defined underlying theory. 
\item
There are potential implications for cosmology and supersymmetry.
\item
A scalar singlet is enough~\cite{eliasmiro} to cure the instability if needed for the consistency of model building.
\end{itemize}

At this point, I would like to open a parenthesis and pay a twofold tribute to SM theory. A theoretical construction that was essentially completed 40 years ago found recently its triumph and stands as solid as a rock: this triggers my deepest admiration for the vision and the insight  of the founding fathers. 
However, I think that today we should also acknowledge the efforts of less glamorous but also very important theorists, who worked for years and years to characterise direct and indirect signals of the SM scalar boson and compute the relevant backgrounds.

Closing the parenthesis, let me move to another key question: are we done now? 
The answer is of course negative: the 2012 discovery and the first studies of The Boson's properties presented at this meeting are just the start of a major program, which may take several decades to  reach completion. 
The two main points of the program for the years to come are obvious:
\begin{enumerate}
\item
Study the properties of The Boson with the highest possible precision, to reveal possible inconsistencies of the SM that would point indirectly to new physics.
\item
Find out whether The Boson is accompanied by other new physics near the TeV scale. 
\end{enumerate}
Both missions may require an electron-positron collider to complement the unique information that the LHC will collect until about 2030.
The general strategy for the years to come was reviewed by Jenni~\cite{jenni}, who pointed out the complementarity between the high-luminosity phase of the LHC (HL-LHC) and possible future high-energy electron-positron machines for precision studies of the properties of The Boson. 

\begin{figure}[bth]
\begin{minipage}{0.48\linewidth}
\centerline{\includegraphics[width=0.9\linewidth]{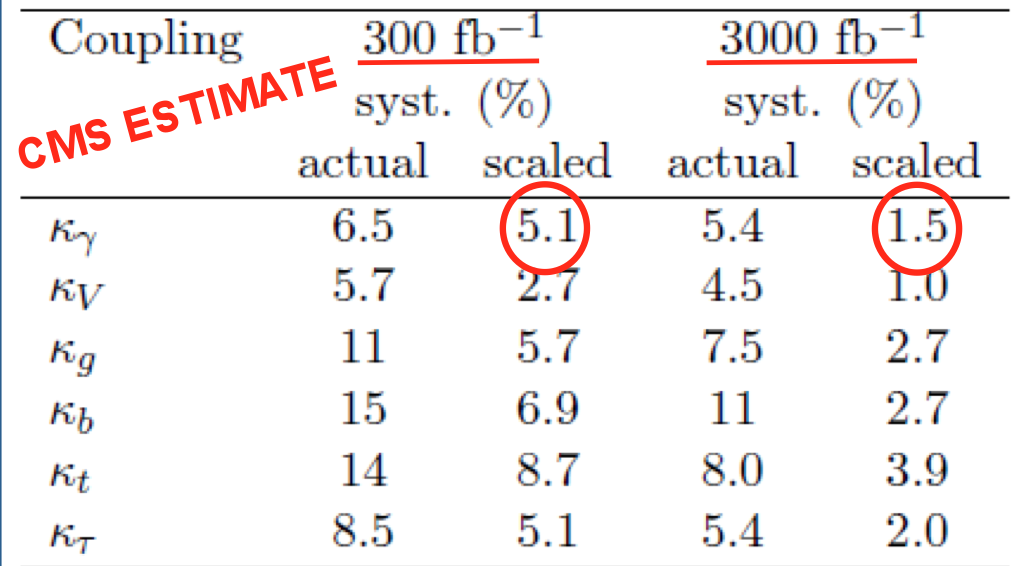}}
\end{minipage}
\hfill
\begin{minipage}{0.48\linewidth}
\centerline{\includegraphics[width=0.9\linewidth]{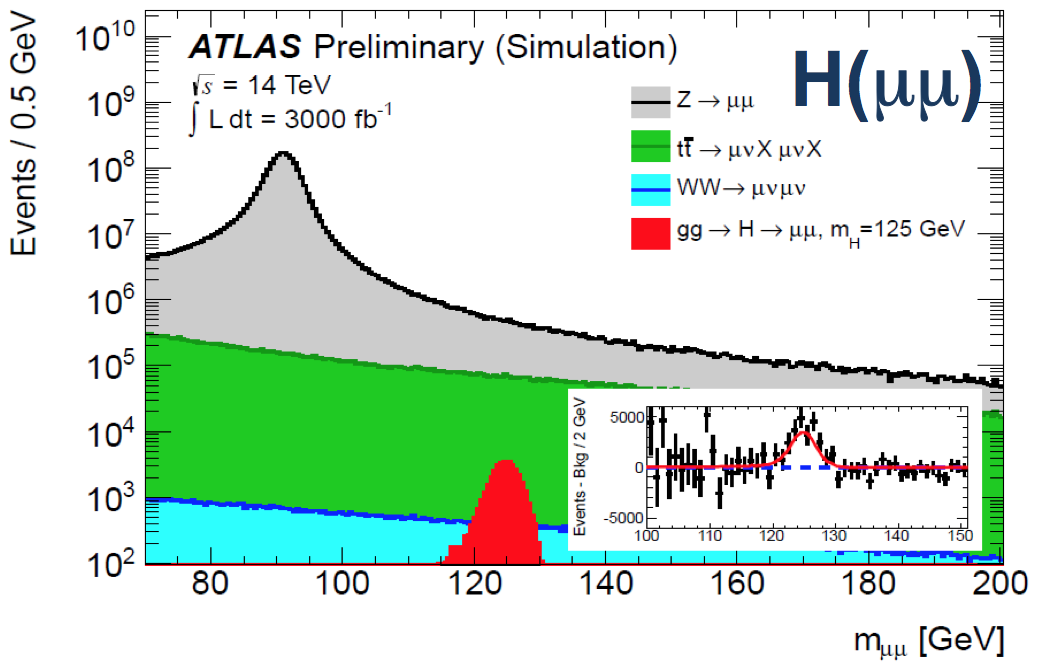}}
\end{minipage}
\caption{Some preliminary estimates of the reach of HL-LHC for the study of the Higgs boson couplings.
Left: the expected precision attainable by CMS with 300 fb$^{-1}$ and 3000 fb$^{-1}$,  keeping the present systematic errors or scaling both experimental and theoretical errors according to some reasonable assumptions. Right: the expected invariant mass distributions for 3000 fb$^{-1}$ at $\sqrt{S} = 14$~TeV , for inclusive $H \rightarrow \mu \mu$ in ATLAS.}
\label{fig:HLLHC}
\end{figure}

Some representative examples of the reach of HL-LHC are shown~\cite{jenni} in Fig.~\ref{fig:HLLHC}: on the left-hand side, the determination of the effective couplings  ($\kappa_\gamma,\kappa_V, \kappa_g, \kappa_b, \kappa_t, \kappa_\tau$), assuming that there are no additional BSM contributions to the total Higgs width; on the right-hand side, the expected sensitivity to inclusive $H \rightarrow \mu \mu$, which could exceed 5$\sigma$.

\begin{figure}[htb]
\centerline{\includegraphics[width=0.5\linewidth]{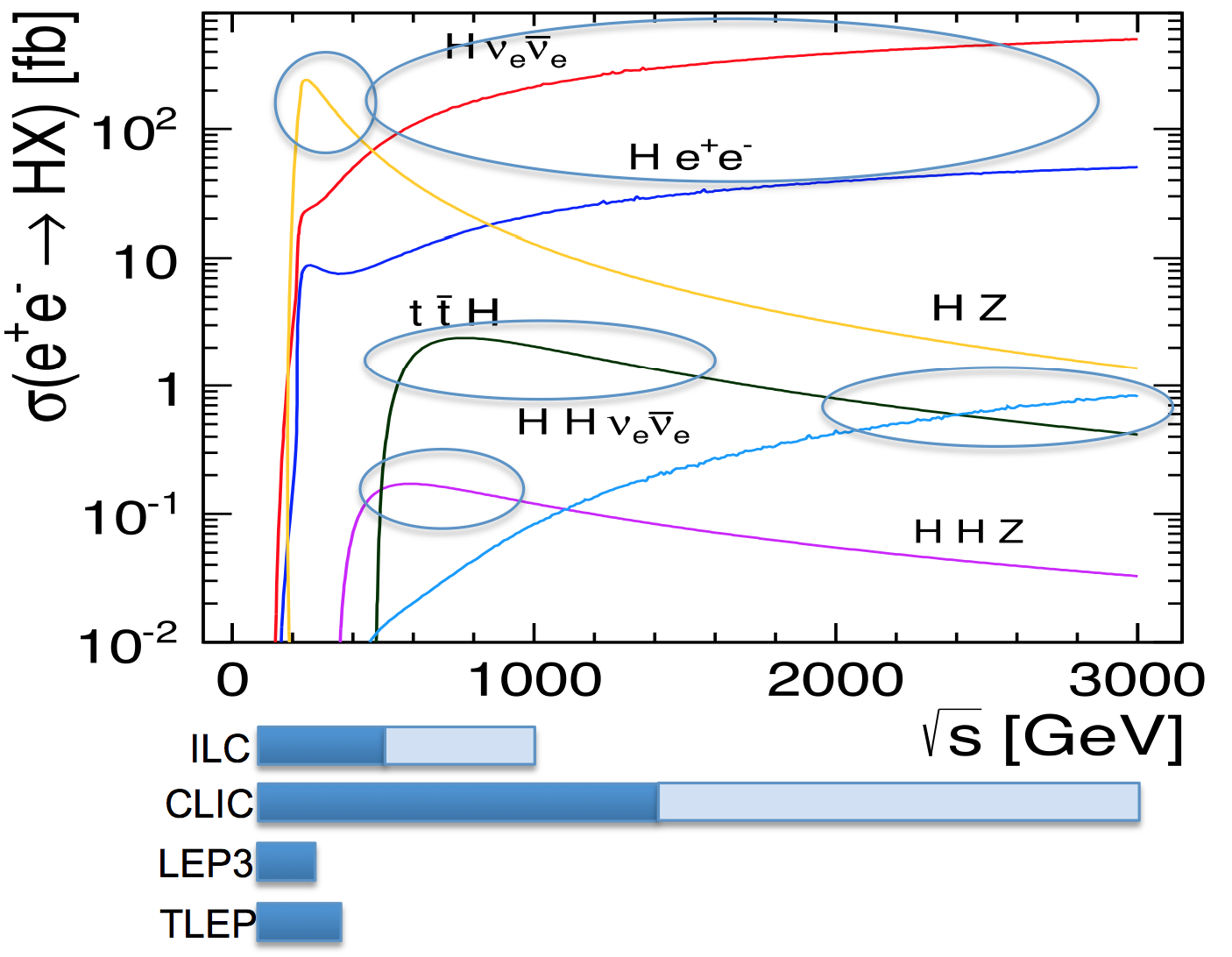}}
\caption{Cross-sections for some of the most important Higgs production processes at $e^+e^-$ colliders, as functions of the centre-of-mass energy $\sqrt{s}$. The design energies of some proposed machines and  their upgrades are also shown.}
\label{fig:strategy}
\end{figure}

The cross-sections for various $H$ production processes in $e^+e^-$ collisions are shown~\cite{jenni} in Fig.~\ref{fig:strategy}. 
The crucial feature is the possibility of directly reconstructing the $HZZ$ coupling from the process $e^+ e^- \rightarrow H Z$ in a model-independent way, exploiting the recoil mass against a leptonically decaying $Z$ boson.
However, energy is also an important factor if we want to test the cubic self-coupling via double $H$ production or the direct coupling to the top quark via associated production with $t \overline{t}$.


\section{Beyond the Standard Model}

Leaving the SM-like scalar particle aside, for which the desirable future experimental program is quite clear, the recent LHC results do not help us much in answering a really difficult question: what lies Beyond the SM, with a chance of being at reach?

The first run of the LHC was rewarded by a historical discovery, but it relied on a very powerful no-lose theorem, guaranteeing either the SM scalar or New Physics at the TeV scale. 
To avoid fooling ourselves, we must admit that we will not be again in such a condition for a long time. 
We should then diversify efforts, with some judgement, to maximise our chances of further important discoveries.
We are now sailing in uncharted waters, and  we must be persistent in our search for New Physics as Columbus in his trip to Indies. 
And in the end, perhaps, we will find the Americas rather than the Indies \ldots

The usual framework to discuss BSM physics is to look at the SM as an {\em effective field theory}, valid up to some  physical cut-off scale $\Lambda$, and to write down the most general local Lagrangian built out of the SM fields (including the scalar doublet $\phi$) and compatible with the SM symmetries, scaling all dimensionful couplings by appropriate powers of $\Lambda$.
The resulting dimensionless coefficients are then parameters, which can be either fitted to experimental data or (if we are able to do so) theoretically determined from the fundamental theory replacing the SM at the scale $\Lambda$. 
Very schematically, and omitting all coefficients, indices and conjugations:
\begin{eqnarray}
{\cal L}_{eff} & = &
\Lambda^4 + \Lambda^2 \phi^2 
\nonumber \\ & + &
\left( D \phi \right)^2 + \overline{\Psi} \not{\! \! D}
\Psi + F^{\mu \nu} F_{\mu \nu} + F^{\mu \nu} 
\widetilde{F}_{\mu \nu} + \overline{\Psi} \Psi \phi + \phi^4 
\nonumber \\ & +  &
{\overline{\Psi} \Psi \phi^2 \over \Lambda}
+
{\overline{\Psi} \Psi \overline{\Psi} \Psi \over \Lambda^2}
+
{ \phi^2 F^{\mu \nu} F_{\mu \nu} \over \Lambda^2}
+
\ldots \, ,
\label{leff}
\end{eqnarray}
where $\Psi$ stands for the generic quark or lepton field, $F$ for the field strength of the SM  gauge fields, and $D$ for the gauge-covariant derivative.  
The first line of Eq.~(\ref{leff}) contains two operators carrying positive powers of $\Lambda$, a cosmological constant term, proportional to $\Lambda^4$, and a scalar mass term, proportional to $\Lambda^2$. 
Barring for the moment the discussion of the cosmological constant term, which becomes relevant only 
when the model is coupled to gravity, it is important to observe  that {\em no} quantum SM symmetry is recovered by setting to zero the coefficient of the scalar mass term. 
On the contrary, the SM gauge invariance forbids fermion mass terms of the form $\Lambda 
\overline{\Psi} \Psi$. 
The second line of Eq.~(\ref{leff})  contains operators with no power-like dependence on $\Lambda$,
but only a milder, logarithmic dependence, due to infrared renormalization effects between the cut-off scale $\Lambda$ and the electroweak scale. 
The operators of dimension $d \le 4$ exhibit two remarkable properties: all those allowed by the
symmetries are actually present in the SM; both baryon number and the individual lepton numbers are automatically conserved.
The third line of Eq.~(\ref{leff}) is the starting point of an expansion in inverse powers of $\Lambda$, containing infinitely many terms. 
For energies and field values much smaller than $\Lambda$, the effects of these operators are suppressed, and the physically most interesting ones are those that violate  some accidental symmetries of the $d \le 4$ operators. 
For example, it is well known that $d=5$ operators of the form $\overline{\Psi}  \Psi \phi^2/\Lambda$ can generate Majorana neutrino masses  of order $G_F^{-1} / \Lambda$ (where $G_F^{-1/2} \simeq 250$~GeV is the Fermi scale); some of the $d=6$ four-fermion  operators can be associated with flavour-changing neutral currents or with baryon- and lepton-number-violating processes such as proton decay, and so on.

\subsection{BSM with neutrinos}

\begin{figure}
\begin{minipage}{0.54\linewidth}
\centerline{\includegraphics[width=0.9\linewidth]{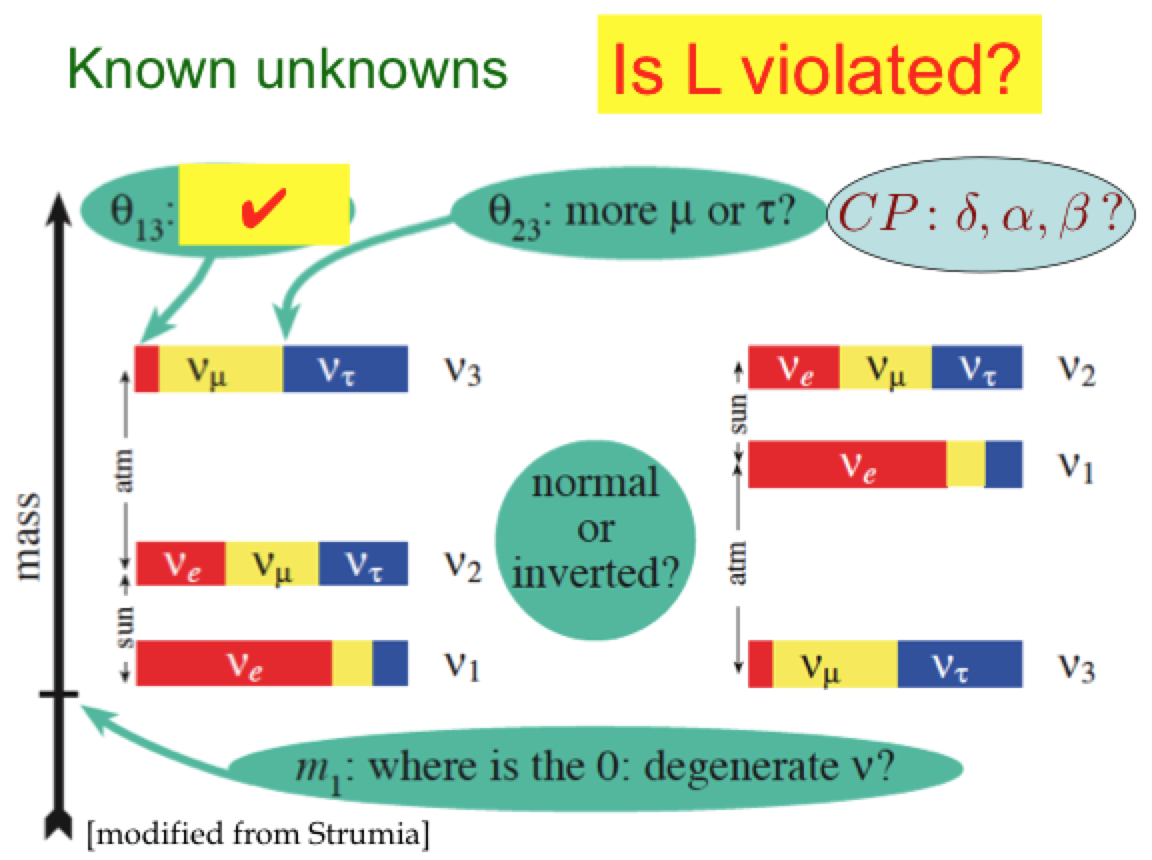}}
\end{minipage}
\hfill
\begin{minipage}{0.46\linewidth}
\centerline{\includegraphics[width=0.9\linewidth]{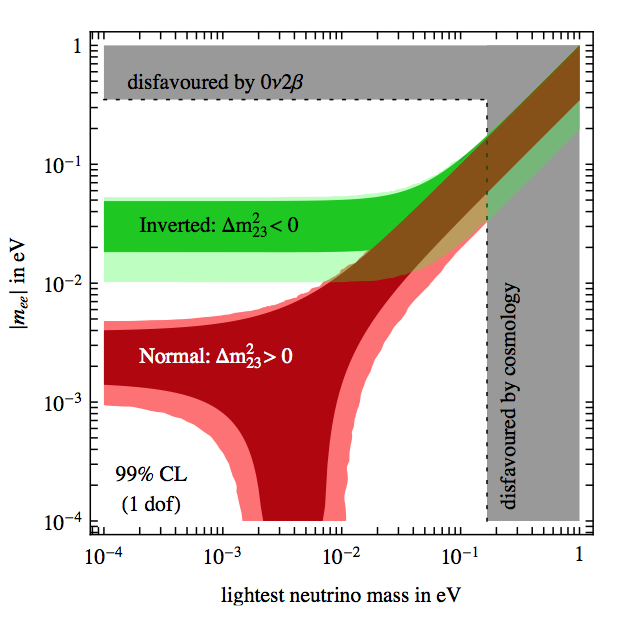}}
\end{minipage}
\caption{Left: a pictorial representation of the most important open questions in neutrino physics. Right: the $90\%$~CL range for the neutrino mass parameter $|m_{ee}|$ controlling neutrinoless double beta decay, as function of the lightest neutrino mass, for inverted (green) and normal (red) hierarchy; the darker regions show how the $|m_{ee}|$ range would shrink if the present best-fit values of oscillation parameters were confirmed with negligible error.}
\label{fig:neutrinos}
\end{figure}

To incorporate neutrino masses and oscillations, which were an important part of the experimental program of this meeting, we must go beyond the minimal SM, and write for the effective Lagrangian valid at the presently explored energy scales something like
\begin{equation}
{\cal L}_{eff} =  {\cal L}_{SM}  + \delta {\cal L} (m_\nu) + \ldots \, , 
\end{equation}
where ${\cal L}_{SM}$ is the minimal, renormalizable SM Lagrangian containing three fermion families, each one with a left-handed neutrino $\nu_L$ but no right-handed neutrino $\nu_R$. 
As already stressed before, ${\cal L}_{SM}$ has an accidental $[U(1)]^4$ global symmetry, associated with the baryon number $B$ and the three partial lepton numbers $(L_e,L_\mu,L_\tau)$, although the combination $(B+L)$ is anomalous and broken by non-perturbative effects. 
The modification $\delta {\cal L} (m_\nu)$ is experimentally needed, but its precise form is still undetermined. 
As is well known, the simplest solutions fall in two classes:
\begin{enumerate}
\item
Neutrinos have Dirac masses: we add a right-handed neutrino, singlet under the SM gauge group, to each fermion family, and assume that $(B-L)$ is a non-anomalous global symmetry of the renormalizable part of ${\cal L}_{eff}$.
\item
Neutrinos have Majorana masses: if we keep the minimal SM symmetries and field content,  there is a unique type of $d=5$ operator in ${\cal L }_{eff}$, explicitly breaking $(B-L)$, and it is precisely of the form required to generate Majorana neutrino masses.
\end{enumerate}
Although there is a clear theoretical bias for option 2,  the choice between options 1 and 2 is in my opinion the most important open question in neutrino physics.  
However, both options are very well compatible with the success of ${\cal L}_{SM}$ in describing all the rest of particle physics up to very high cutoff scales $\Lambda$. 
But there are other important experimental questions in neutrino physics waiting for an answer, which in some cases may be closer in time than expected some years ago, because we have recently learnt that the mixing angle $\theta_{13}$ is non-vanishing and relatively large.
A pictorial representation of the most important open questions in neutrino physics can be found on the left-hand side of Fig.~\ref{fig:neutrinos}~\cite{fsv}. 
There are now good hopes of determining the mass hierarchy and establishing CP violation in the neutrino sector with the next generation of experiments. 
The right-hand side of Fig.~\ref{fig:neutrinos}~\cite{fsv} is there to remind us that, if by any chance an inverted neutrino mass hierarchy were to be experimentally established in the coming decade, neutrinoless double-$\beta$ decay could be at reach for the following generation of experiments, and it would be worth multiplying the efforts towards solving the most important problem of neutrino physics. 

The main issues touched in the theory talks relating to neutrino physics~\cite{gavela}$^, \,$\cite{hernandez}$^, \,$\cite{tamborra}$^, \,$\cite{palomaresruiz} can be concisely summarised as follows: 
\begin{itemize}
\item
Some experimental anomalies (LNSD, MiniBoone, reactors) hint at the possibility of sterile neutrinos with masses in the eV range, although their individual significance is not very high and no consistent picture seems to emerge from the fits~\cite{fits}. As discussed by Tamborra~\cite{tamborra}, there is also a variety of cosmological constraints/hints. My personal opinion is in line with the one expressed in the Experimental Summary: it is difficult to draw firm conclusions, and the new data coming from the Planck mission are unlikely to change significantly the situation concerning this particular issue.
\item
Are we in a better position to develop models of flavour, now that we have information on two flavour sectors, the hadronic and the leptonic one? 
Two interesting approaches were described at this meeting. 
The first one, pursued by several groups over the last decades and discussed by Hernandez~\cite{hernandez}, is based on non-Abelian discrete flavour symmetries.
The second one, discussed by Gavela~\cite{gavela}, tries to revamp the old idea of treating the Yukawa couplings as dynamical variables. 
\item
Can the scale of lepton flavour be low enough to give detectable signals in charged lepton flavour violation? The answer is obviously yes, and justifies pushing further the searches for charged lepton flavour violation, even if there is no compelling argument to be confident in observable deviations from the SM predictions. 
\item
How will we establish the neutrino mass hierarchy? With accelerator-driven long-baseline oscillation experiments, with reactor-driven oscillation experiments with 50-60~km baselines or, as discussed by  Palomares Ruiz~\cite{palomaresruiz}, with multi-Mton extensions of current neutrino telescopes looking at atmospheric neutrinos? Time will tell, it might be a close race as for the determination of $\theta_{13}$ \ldots
\end{itemize}

\subsection{BSM with Dark Matter}

Weakly Interacting Massive Particles (WIMPs) are presently considered among the leading candidates for Dark Matter (DM) because of the so-called WIMP Miracle: for a WIMP in thermal equilibrium after inflation, 
\begin{equation}
\label{eq:wimp}
\langle \sigma_{ann} \, v \rangle \simeq 3 \times 10^{-26} \ {
\rm cm^3 \, s^{-1}} \, , 
\end{equation}
and this is roughly the size of an electroweak cross-section for a particle with mass $M \sim 10^{2-3} \,  {\rm GeV}$.  This is certainly a good argument for having New Physics at TeV scale, but not a fully compelling one: DM could well be axions.
\begin{figure}
\centerline{\includegraphics[width=0.5\linewidth]{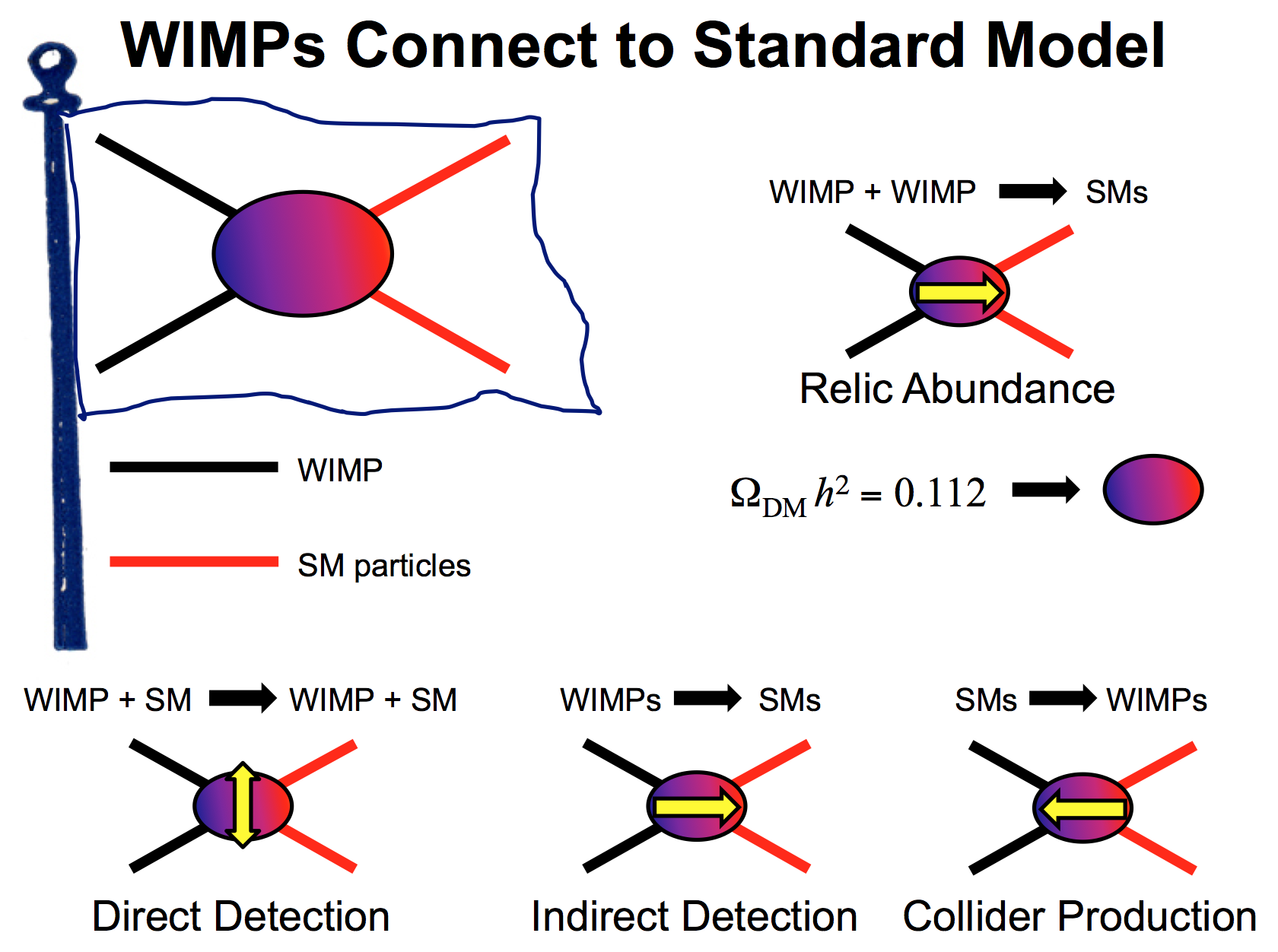}}
\caption{Different processes involving WIMPs and SM particles in their initial and final states.}
\label{fig:DM}
\end{figure}
What makes WIMPs very attractive candidates for DM is, besides their possible connection with the naturalness problem of the SM (see subsection 3.3), the multiplicity of ways in which they could give rise to experimentally detectable signals, as sketched~\cite{kolb} in Fig.~\ref{fig:DM}.
By switching some of the external legs from the initial to the final state, or viceversa, the same Feynman diagrams can describe: DM annihilation into SM particles, which controls the relic abundance and can also give rise to indirect signals of DM in cosmic rays; direct DM detection through elastic scattering on targets made of SM particles; finally, annihilation of SM particles into final states containing DM particles, which could give rise to detectable signals at the LHC or other high-energy colliders.

Several theoretical issues connected with DM were discussed at this meeting~\cite{ibarra}$^, \,$\cite{rydbeck}$^, \,$\cite{tytgat}$^, \,$\cite{lopezhonorez}: the talks explored a variety of models that could produce diverse and interesting signals in indirect or in collider searches. 
Without going into details, I would like to make two general comments:
\begin{enumerate}
\item
The improved LHC bounds on new particles coming from direct searches, which particularly affect those particles coupling significantly to quarks and gluons, may eventually shift the focus on WIMP candidates from "social" ones (such as the LSP in supersymmetric extensions of the SM), to "simplified" DM models.
\item
It is more and more frequent to see comparisons between the reach of collider searches for DM, for example those looking for monojets or single photons at the LHC, and direct searches for DM in low-energy experiments, which make use of effective Lagrangians containing local four-particle operators with two DM particles and two SM particles. When making such comparisons, it is very important to check that they fall within the region of validity of the underlying theoretical approximation, which is not always the case. 
\end{enumerate}

\subsection{Naturalness}

Naturalness~\cite{naturalness}$^, \,$\cite{hooft} has been one of the guiding principles of BSM theory in the last decades, and in particular the main reason to expect BSM physics at the TeV scale. A synthetic but powerful formulation of the concept of naturalness is due to 't~Hooft~\cite{hooft}: in an effective Lagrangian such as the one of Eq.~(\ref{leff}), a dimensionless coefficient in front of a local operator, weighted by the appropriate power of $\Lambda$, can be small only if we recover an exact quantum symmetry in the limit in which the coefficient goes to zero.
We know of a number of cases in which the concept of naturalness is successfully at work:
\begin{enumerate}
\item
The electron mass $m_e$ in classical electrodynamics (see e.g. \cite{murayama}): we would naively compute linearly divergent corrections, $\delta m_e \sim \alpha \, \Lambda$, and the New Physics rescuing naturalness is the positron, with mass $m_e$.
\item
The four-fermion operators describing strangeness-changing neutral current processes in the Fermi theory with three light quarks.
The quadratically divergent one-loop contribution, going as $G_F^2 \, \Lambda^2$, must be cut off at the GeV scale \cite{kaons}, and the New Physics coming to rescue naturalness is the charm quark~\cite{maiani}.
\item
The mass difference between the charged and the neutral pion in scalar QED~\cite{pions}.
At one loop we would obtain $\Delta m_\pi^2 = m^2 (\pi^+) - m^2(\pi^0) \sim (3 \alpha)/(4 \pi) \, \Lambda^2$, which calls for new physics near the GeV to preserve naturalness: such new physics in this case is the  $\rho$ meson.
\end{enumerate}
However, we also know of a very important example where the criterion of naturalness seems to fail, at least according to our present understanding. 
It is the Dark Energy problem, or the cosmological constant problem, whose naturalness scale in particle physics units would be $\Lambda_{CC} \sim 2.4 \times 10^{-3} \, {\rm eV}$. 
Are we missing some subtle modification of the gravitational interactions involving new degrees of freedom at such a mass scale? 
Tests of the gravitational interactions at short distances now extend well below the millimeter scale and do not leave much room. 
Is there a suitable realization of conformal symmetry at the quantum level that, after a delicate enough breaking, can account simultaneously for the two hierarchies $M_{weak}/M_P \sim 10^{-15}$ and $\Lambda_{CC}/M_P \sim M_{weak}^2/M_P^2 \sim   10^{-30}$? 
Some attempts along this direction did not go very far (see e.g. \cite{bardeen}$^, \,$\cite{locconf} and the papers quoting them), but the difficulty of more conventional approaches to naturalness is encouraging many to revisit the problem, to see whether something important has been missed and can now be identified.   
 
In the framework of the SM, the naturalness problem boils down to the fact that no exact quantum SM symmetry is recovered in the limit $M_H \rightarrow 0$. 
Classically, we would recover invariance under scale transformations, but in the SM such an invariance is broken by quantum corrections and, even more importantly, it is difficult to imagine a scale-invariant ultraviolet theory underlying the SM. 
This is what has led many to declare that the SM is unnatural unless there is New Physics at the LHC scale.
Since, within the SM, the leading 1-loop quadratically divergent corrections to the scalar mass parameter come from the top quark loop, and go as $\delta M_H^2 \sim - (3 \, h_t^2)/(8 \pi^2) \, \Lambda^2$, requiring $\delta M_H^2 < {\cal O} (M_H^2)$ roughly requires $\Lambda < {\cal O} (500) \ {\rm GeV}$. 
However, precision tests of the SM suggest $\Lambda > {\cal O} (few) \ {\rm TeV}$ for generic flavour-independent operators, and $\Lambda > {\cal O} (1000) \ {\rm TeV}$ for generic flavour-dependent operators.
Moreover, no BSM particle has been found at the LHC to date~\cite{sphicas}.
What is the way out?

The most conservative attitude is to insist on the few almost-natural models that survive the present experimental data, where \emph{almost-natural} means that we leave room for some accidental cancellation at the few-percent level. 
The two leading classes of models along these lines are a quite restricted subset of supersymmetric extensions of the SM and of models in which the SM scalar is a (partially) composite state generated by a new strong interaction. 
The important fact to stress here is that all these models predict new particles within the reach of the (full-energy) LHC, therefore they can all be ruled out at the end of the LHC experimental program. 

Have we missed some more subtle possibilities, perhaps in connection with gravity and Dark Energy?
Of course, the logical possibility remains that the puzzle might be solved only in the full theory, by a mysterious infrared-ultraviolet connection missed by the effective field theory.

In the context of the discussions on naturalness, an important subject is how natural extensions of the SM can be reconciled with precision tests of the SM in the flavour sector. 
In other words, the excellent agreement between the experimental data and the SM description of flavour  breaking  would suggest a scale of new physics $\Lambda > 10^3$--$10^4$~TeV, whereas naturalness would call for a scale of new physics $\Lambda < {\cal O} (500) \, {\rm GeV}$. 
A possible way out is to make the dimensionless coefficients in front of the flavour-breaking higher-dimensional operators, for example those of dimension 6, much smaller than one because of some flavour symmetry. 
An interesting possibility discussed at this meeting~\cite{sala} is to consider models based on a $U(2)^3 = U(2)_{Q_L} \times U(2)_{U_R} \times U(2)_{D_R}$ flavour symmetry in the quark sector: in such a case, the dimension-six effective operators with potentially dangerous flavour-breaking structures can receive additional suppressions of order $V_{CKM}^{2-4}$, and this can still reconcile the stringent bounds from flavour physics with a cutoff scale $\Lambda$ of a few TeV, with a potentially interesting phenomenology just waiting to be discovered.   

\begin{figure}
\begin{minipage}{0.59\linewidth}
\centerline{\includegraphics[width=0.9\linewidth]{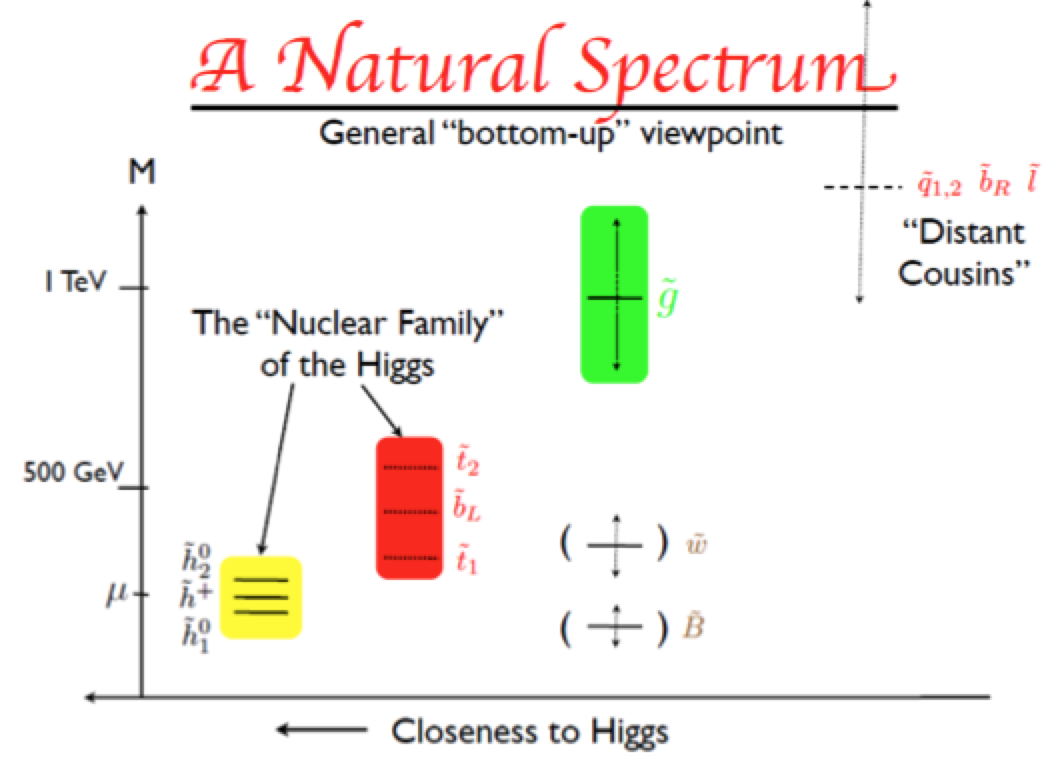}}
\end{minipage}
\hfill
\begin{minipage}{0.39\linewidth}
\centerline{\includegraphics[width=0.9\linewidth]{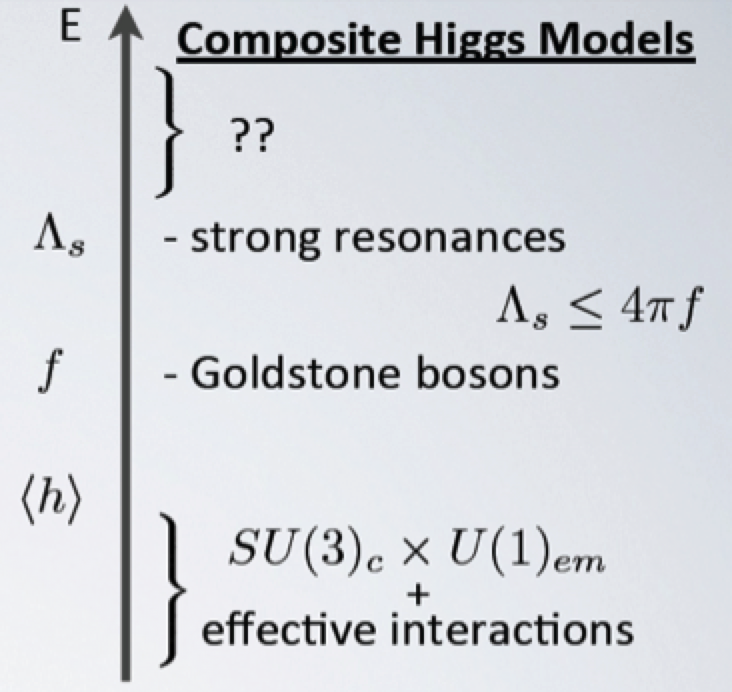}}
\end{minipage}
\caption{Surviving almost-natural models: minimal SUSY and partial compositeness}
\label{fig:natural}
\end{figure}

In the context of supersymmetric extensions of the SM, one out of several attempts at reconciling naturalness with the new stringent LHC bounds is to consider minimal models, in which only the supersymmetric particles most closely coupled to the Higgs field are bound to be light. A typical spectrum is shown~\cite{hall} on the left-hand side of Fig.~\ref{fig:natural}.
This possibility is receiving increasing attention in the experimental searches at the LHC~\cite{marrouche}$^,$\cite{verducci}, and the results presented at this meeting show that the available parameter space has already started to shrink considerably, although unexplored regions remain.  

As for the models with an almost-natural composite scalar, their typical structure is shown~\cite{merlo} on the right-hand side of Fig.~\ref{fig:natural}. 
The talks by Merlo~\cite{merlo} and Azatov~\cite{azatov} emphasized the correlation between a light partially composite scalar and light fermionic partners of the top quark. 
The talk by Yamawaki~\cite{yamawaki} reminded us that walking technicolor with approximate scale invariance may allow for a light technidilaton, which could mimic The Boson found at the LHC. However, significant deviations from the SM couplings are predicted, and my impression is that the new data presented at this meeting might be already sufficient to strongly disfavour this possibility. 
 
\subsection{Supersymmetry and ascendants}

The status of supersymmetric extensions of the SM, from phenomenological models at the TeV scale to possible string-motivated ascendants, was reviewed at the meeting in several talks~\cite{kazakov}$^, \,$\cite{carena}$^, \,$\cite{ellwanger}$^, \,$\cite{dudas}$^, \,$\cite{sagnotti}. Again, I will not review them in detail but rather transmit you, perhaps in an oversimplified way because of time constraints, my present opinion, after following the experimental searches and contributing to the theory for roughly thirty years:
\begin{itemize}
\item
Supersymmetry is too good an idea to be wasted by Nature: its role as a general symmetry of Relativistic Quantum Field Theory, the very existence of supergravity as the gauge theory of supersymmetry and the role it plays in superstrings, are sufficient for me to consider supersymmetry as a plausible ingredient of whatever fundamental theory underlies the SM. But at what physical scale? 
\item
Supersymmetry, whose main phenomenological motivation has been for decades the resolution of the SM naturalness problem, is now in trouble with naturalness, at least in its simplest realizations. Attempts at formulating almost-natural supersymmetric models were mentioned in the previous subsection.  Another logical possibility is that supersymmetry might need to be combined with some (not yet identified) additional ingredient to solve the SM naturalness problem.
\item
Conventional supersymmetric models (CMSSM, NMSSM, \ldots) do not work as such and should finally rest in peace, although some of their features may contain some truth: for example, the additional contributions to $M_H$ present in the NMSSM with respect to the MSSM may play a role in improving naturalness. But today it is difficult for me to believe that the CMSSM or the NMSSM could be correct in all their details.
\end{itemize}

Concerning the present status of SUSY phenomenology, I feel like quoting a sentence from another summary talk, given by H.~Georgi at a conference in Santa Barbara I attended in the early 90Õs (giving a talk on supersymmetric phenomenology): \emph{`Stop wandering in susy parameter space!'} 
At the moment I was not very happy to hear it, but in retrospect it was a wise suggestion: as theorists we can be more useful for the progress of the field than by making computer-assisted scans of the parameter spaces of supersymmetric models with fancy statistical tools. 
For example, if we want to be close to experiment, we can point out possible signals that may arise in plausible models, and that our experimental colleagues  may have overlooked so far in their searches. 
Alternatively, we could try to understand what theoretical ingredient we have been missing in our approach to naturalness within simple, controllable, non-realistic contexts, and come back to realistic models and to our experimental colleagues only after making some significant progress.  

Of course, supersymmetry could be relevant to our description of Nature and perhaps accessible to the LHC even if it fails to address the naturalness problem.  Indeed, if we do not insist on naturalness, SUSY  with heavy scalars can evade direct searches  and flavour constraints, while mantaining gauge coupling unification and an acceptable DM candidate. An example, reviewed in the talk by Dudas~\cite{dudas}, is mini-split supersymmetry: a typical spectrum could include  scalars at ~1000 TeV, gauginos 1 or 2 loop factors lighter (protected by an R-symmmetry), higgsinos with model-dependent masses, a SM-like scalar with mass $M_H \simeq 125 \ {\rm GeV}$ easily  reproduced.

\section{Conclusions}

In conclusion, and waiting for new results at the Summer Conferences, the main messages I can extract from this Winter 2013 Electroweak Moriond are those summarised in Table~1: 
\begin{itemize}
\item
Experimental particle physics has entered a phase of accelerated progress.
\item
The SM is still untarnished after crossing new thresholds in energy and precision, and its triumph is going beyond most expectations.
\item
We have many reasons to believe in BSM physics, but we do not know for sure where to look for it, thus we must keep an open mind and diversify our efforts.
\item
The concept of naturalness is being seriously challenged: whatever the result of the 13-14 TeV run of the LHC will be, this will have a profound impact on our theoretical approach to the fundamental interactions.
\end{itemize}

Personally, I am convinced that we have just started to digest the implications of the experimental results presented at this meeting: to fully exploit all the new information and to prepare for the more experimental data to come, a lot of theoretical work is still required, in several directions. 
We must know, we will know, but we must be patient and try hard: we are lucky to live in such exciting times!

As the last speaker, I would like to conclude by warmly thanking, on behalf of all the participants, all those who contributed with their hard work to the success of this meeting:  the Secretariat (Isabelle Cossin, Vera de Sa-Varanda, Sarodia Vydelingum), the Computer Support (G\'erard Dreneau, Damien Fligiel, Victor Mendoza, Olivier Drevon),  the Organizers (Etienne Aug\'e, Jacques Dumarchez,  Jean Tran Thanh Van),  the Scientific Organizers present in La Thuile (Lydia Iconomidou-Fayard, Jean-Marie Fr\`ere,  Eric Armengaud, Patrick Janot, Jean-Pierre Lees, Sotiris Loucatos, Francois Montanet, Jean Orloff), as well as all the other Scientific Organisers who contributed remotely.

\section*{Acknowledgments}
This work was supported by the ERC Advanced Grant No.267985 \emph{Electroweak Symmetry Breaking, Flavour and Dark Matter: One Solution for Three Mysteries} (\emph{DaMeSyFla}), by the European Programme \emph{UNILHC} (contract PITN-GA-2009-237920) and by the MIUR FIRB-2012 Grant RBFR12H1MW \emph{Una nuova forza, l'origine della massa e LHC}.
% I would also like to thank \ldots for useful discussions.

\section*{References}
\end{document}